\documentclass[conference]{IEEEtran}
\IEEEoverridecommandlockouts

\usepackage{graphicx}
\usepackage{subcaption}
\usepackage{cite}
\usepackage{amsmath,amssymb,amsfonts}
\usepackage{algorithmic}
\usepackage{graphicx}
\usepackage{textcomp}
\usepackage{xcolor}

\def\BibTeX{{\rm B\kern-.05em{\sc i\kern-.025em b}\kern-.08em
    T\kern-.1667em\lower.7ex\hbox{E}\kern-.125emX}}

\begin{document}

\title{SplatStream: Fine Granular Scalable Gaussian Splatting for Adaptive 3D Scene Streaming}

\author{\centering
\IEEEauthorblockN{Muhammad Talha}
\IEEEauthorblockA{
\textit{University of Missouri--Kansas City}\\
mtgcf@umkc.edu}
\and
\IEEEauthorblockN{William Gordon}
\IEEEauthorblockA{
\textit{BASIS Independent Silicon Valley}\\
william-gordon@outlook.com
}
\and
\IEEEauthorblockN{Sajid Umair}
\IEEEauthorblockA{
\textit{University of Missouri--Kansas City}\\
suzqnu@umkc.edu}
\and

\IEEEauthorblockN{Zhu Li}
\IEEEauthorblockA{
\textit{University of Missouri--Kansas City}\\
lizhu@umkc.edu}
\and
\IEEEauthorblockN{Anique Akhtar}
\IEEEauthorblockA{
\textit{Qualcomm, Inc.}\\
aniquea@qti.qualcomm.com}
\and
\IEEEauthorblockN{Joel Jung}
\IEEEauthorblockA{
\textit{Qualcomm, Inc.}\\
joeljung@qti.qualcomm.com}
 
}
\vspace{-5mm}

\maketitle

\begin{abstract}
Dynamic 3D Gaussian Splatting (GS) enables high-quality real-time rendering for immersive media, but its large representation size and frame-wise redundancy create significant challenges for adaptive streaming. This paper presents SplatStream, a fine granular scalable Gaussian splatting framework for dynamic 3D scene delivery. The proposed method decompose the GS scenes into quality and resolution layers, and introduces inter-layer predictive coding to achieve scalability. For temporal direction, B-frames are introduced to have temporal quality scalability.  A lightweight cross-layer transformer based predictor is utilized for both cross layer and temporal predictions.  In addition, a volume-opacity based importance measure is used for fine-grained Gaussian packetization, allowing visually important primitives to be transmitted earlier for progressive refinement. Finally, the scalable GS bitstream is mapped to an MPEG-DASH compatible sub-representation structure, enabling fine granular adaptive, low-latency delivery of dynamic Gaussian splatting content under bandwidth-varying conditions.
\end{abstract}


\vspace{-3mm}
\section{Introduction and Motivation}

3D Gaussian Splatting (3DGS) has emerged as an efficient representation for real-time novel-view synthesis, providing high visual quality through rasterization-based rendering~\cite{kerbl2023_3dgs}. By modeling a scene with anisotropic Gaussian primitives carrying geometry, opacity, scale, rotation, and spherical harmonic (SH) appearance attributes, 3DGS enables interactive rendering for immersive applications such as AR/VR, telepresence, and free-viewpoint video. However, high-quality 3DGS scenes often contain hundreds of thousands or millions of Gaussians with high-dimensional attributes, making storage and transmission expensive. This problem becomes more severe for dynamic scenes, where Gaussian frames must be delivered continuously under bandwidth and latency constraints, while also satisfying pixel-rate limits imposed by the number of rendered pixels the decoder must process per second.

Recent 3DGS compression methods reduce storage using pruning, quantization, compact attribute modeling, anchor structures, context modeling, and entropy coding. Compact-3DGS, LightGaussian, EAGLES, Scaffold-GS, CompGS, HAC, and ContextGS show that Gaussian primitives and attributes contain substantial redundancy~\cite{lee2024_compact3dgs,fan2023_lightgaussian,girish2023_eagles,lu2024_scaffoldgs,liu2024_compgs,chen2024_hac,wang2024_contextgs}. In addition, L-GSC provides a lightweight open-source 3D Gaussian Splat codec for encoding and decoding GS representations~\cite{qualcomm2026_lgsc}. 
In parallel, MPEG G-PCC v1 Amd.1 and V-PCC v1 Amd.1 have been used as standardized anchors for Gaussian splatting data, since Gaussian centers and attributes can be organized in point-cloud-like forms and compressed using geometry, attribute, or video-based point-cloud coding tools~\cite{iso2023_gpcc,iso2023_vpcc,schwarz2019_pcc}. While these methods provide strong compression performance, they are mainly designed for compact storage or frame-level coding, and do not fully address adaptive streaming requirements such as fast startup, resolution switching, temporal scalability, and fine-grained partial decoding.

Streaming-oriented 3DGS methods have begun to address these requirements. Fine-granular scalable Gaussian splatting coding introduced an intra-coded streamable framework with progressively decodable Gaussian packets~\cite{zou2026_finegranular}, while low-latency scalable Gaussian splatting coding and LTS further explored DASH-like multi-layer organization for immersive Gaussian scenes~\cite{shi2025_lowlatency,sun2025_lts}. These works are related to MPEG-DASH, where media is organized into representations, segments, and sub-representations for adaptive delivery~\cite{sodagar2011_mpegdash}. However, existing scalable 3DGS streaming approaches are still limited: many are intra-frame oriented, focus mainly on packetization, or do not jointly combine multi-resolution rendering, learned inter-quality-layer prediction, temporal prediction, and fine-grained Gaussian refinement in a unified dynamic streaming framework.

For dynamic 3DGS, temporal redundancy is also critical. Consecutive frames often share similar geometry and appearance despite motion, deformation, visibility changes, and topology variation. InterGS introduced inter-predictive coding for Gaussian splatting sequences, and InterGS-Lite further improved this direction using lightweight prediction and vector quantization of residuals~\cite{talha2025_intergs,talha2026_intergslite}. Nevertheless, these works are predictive coding tools rather than complete scalable streaming systems: they do not define spatial quality layers, temporal enhancement layers, or DASH-compatible sub-representations.

In this work, we propose \textit{SplatStream}, a fine-grained scalable Gaussian splatting framework for adaptive streaming of dynamic 3D scenes. SplatStream treats 3DGS coding as a streamable representation design problem. It uses multi-resolution rendering supervision for scalable intra anchors, introduces a lightweight transformer predictor to reduce redundancy between quality layers, integrates InterGS-Lite for 1080p temporal prediction, and orders Gaussians using a volume-opacity importance criterion inspired by rendering-free primitive importance estimation~\cite{yang2026_rap}. The resulting bitstream supports spatial startup quality, temporal scalability, and fine-grained DASH-style refinement.

The main contributions are:
\begin{itemize}
    \item \textbf{Multi-resolution GS rendering:} We supervise the original-resolution Gaussian representation at multiple rendering resolutions, allowing compact startup at low quality and progressive refinement toward high-resolution playback.
    
    \item \textbf{Transformer-based inter-quality-layer prediction:} A lightweight transformer predicts higher-quality Gaussian attributes from decoded lower-quality layers, reducing redundancy in enhancement-layer coding.
    
    \item \textbf{Temporal scalability with inter-frame prediction:} InterGS-Lite is integrated as the temporal coding module, and dynamic frames are organized into base temporal and B-layer enhancement representations.
    
    \item \textbf{MPEG-DASH sub-representation mapping:} Spatial layers, temporal layers, and volume-opacity refinement units are organized into DASH-compatible representations and sub-representations for adaptive, fine-grained, low-latency streaming.
\end{itemize}

\vspace{-3mm}
\section{Methodology}
\vspace{-2mm}

\subsection{Overview of SplatStream}

SplatStream provides a scalable streaming structure for dynamic 3D Gaussian Splatting (3DGS). Given Gaussian frames $\{\mathcal{G}_{t}\}_{t=1}^{T}$, each primitive is represented as
\vspace{-2mm}
\begin{equation}
    g_i=\{\mathbf{x}_i,\mathbf{s}_i,\mathbf{q}_i,\alpha_i,\mathbf{c}_i\},
    \vspace{-2mm}
\end{equation}
where the terms denote center, scale, rotation, opacity, and SH appearance coefficients, respectively.

SplatStream supports three complementary forms of scalability: spatial-quality scalability, temporal scalability, and fine-grained Gaussian scalability. The first frame of each GOP is treated as an intra anchor. Its Gaussian model is trained at the original scene resolution, but the rendering loss is evaluated at multiple target resolutions, e.g., $540p$, $720p$, and $1080p$, to make the same representation effective for different display qualities. Temporal prediction is performed only in the reconstructed $1080p$ Gaussian domain using an extended InterGS-Lite framework, where a lightweight transformer predictor is added to the original KNN-bilateral predictor set. Finally, opacity-volume Gaussian ordering and MPEG-DASH sub-representations enable fine-grained adaptive streaming.

\vspace{-2mm}
\subsection{Multi-Resolution Intra Rendering Loss}
\vspace{-1mm}
The multi-resolution loss makes the intra anchor suitable for scalable streaming rather than only full-resolution offline rendering. In adaptive delivery, early packets should preserve dominant scene structure for low-rate startup, while later packets refine the reconstruction as bandwidth improves. Importantly, we do not train separate PLY models for $540p$, $720p$, and $1080p$; instead, the same intra Gaussian representation is rendered and supervised at multiple target resolutions.

Let $t=a$ denote the first frame of a GOP and let $\mathcal{R}=\{540,720,1080\}$ be the set of target rendering resolutions. The intra representation $\mathcal{G}_{a}$ is rendered at resolution $r\in\mathcal{R}$ as
\vspace{-2mm}
\begin{equation}
    \hat{\mathbf{I}}_{a}^{r}
    =
    \Phi(\mathcal{G}_{a},r),
    \vspace{-2mm}
\end{equation}
where $\Phi(\cdot)$ denotes the Gaussian rasterizer and $\mathbf{I}_{a}^{r}$ is the ground-truth image downsampled to resolution $r$. The multi-resolution intra rendering loss is
\vspace{-2mm}
\begin{equation}
\begin{aligned}
\mathcal{L}_{\text{MR}}
=
\sum_{r\in\mathcal{R}}
\lambda_r
\Big[
& (1-\beta)
\left\|
\hat{\mathbf{I}}_{a}^{r}
-
\mathbf{I}_{a}^{r}
\right\|_1  \\
&+
\beta
\mathcal{L}_{\text{SSIM}}
\left(
\hat{\mathbf{I}}_{a}^{r},
\mathbf{I}_{a}^{r}
\right)
\Big],
\end{aligned}
\end{equation}
where $\lambda_r$ controls the contribution of each resolution and $\beta$ balances pixel distortion and structural similarity.

After training, scalable intra packets are generated by controlling the transmitted Gaussian subset and enhancement information. Lower-rate packets support compact startup, while higher-rate packets refine the reconstruction toward full quality. The decoded full-quality intra representation then serves as the reference for the $1080p$ temporal prediction chain.

\begin{figure}[t]
    \centering
    \includegraphics[width=\linewidth]{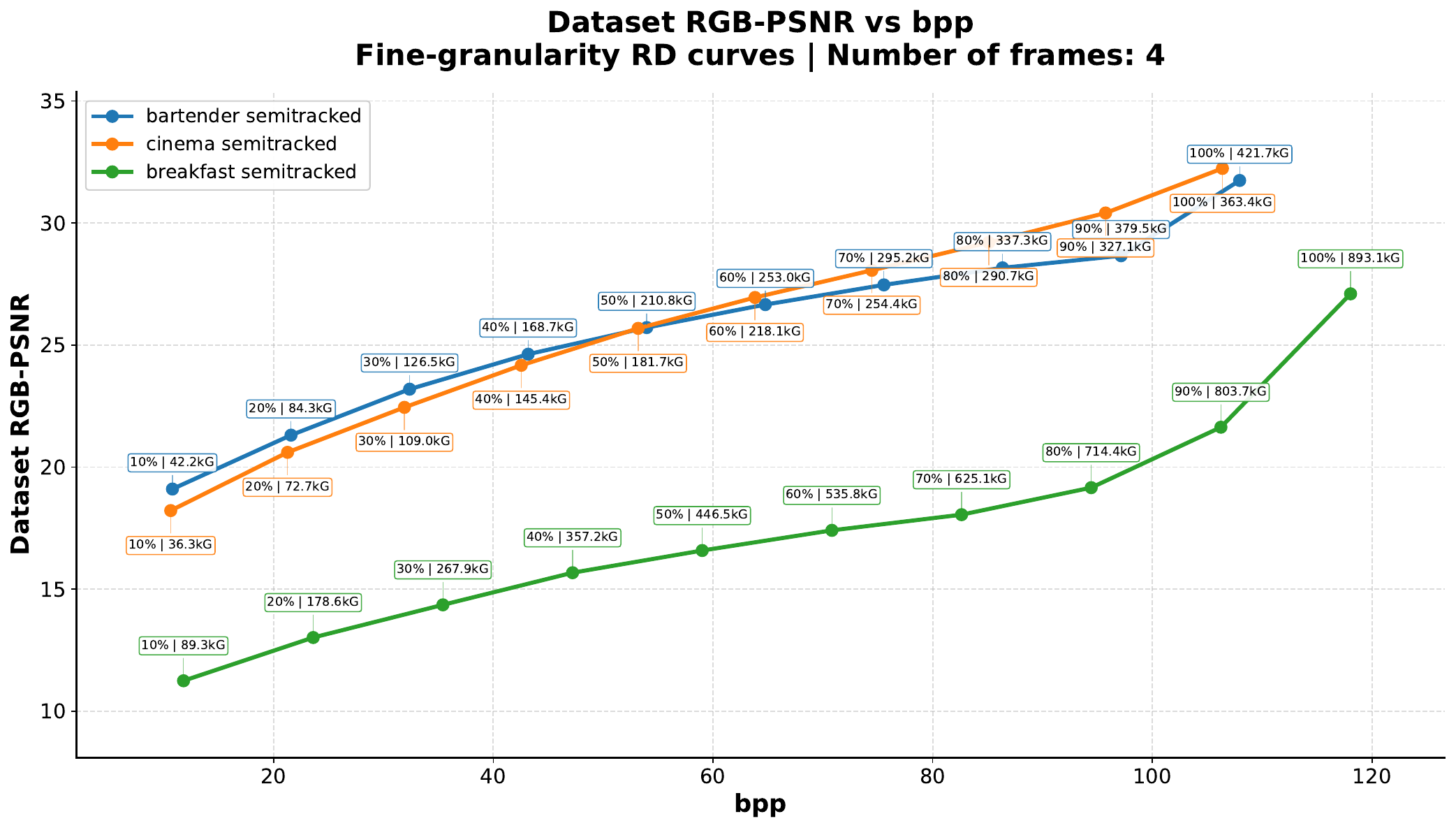}
    \caption{RGB-PSNR versus bpp for fine-grained refinement levels across three dynamic 3DGS sequences.}
    \label{fig:rd_psnr}
    \vspace{-4mm}
\end{figure}

\begin{figure}[t]
    \centering
    \includegraphics[width=\linewidth]{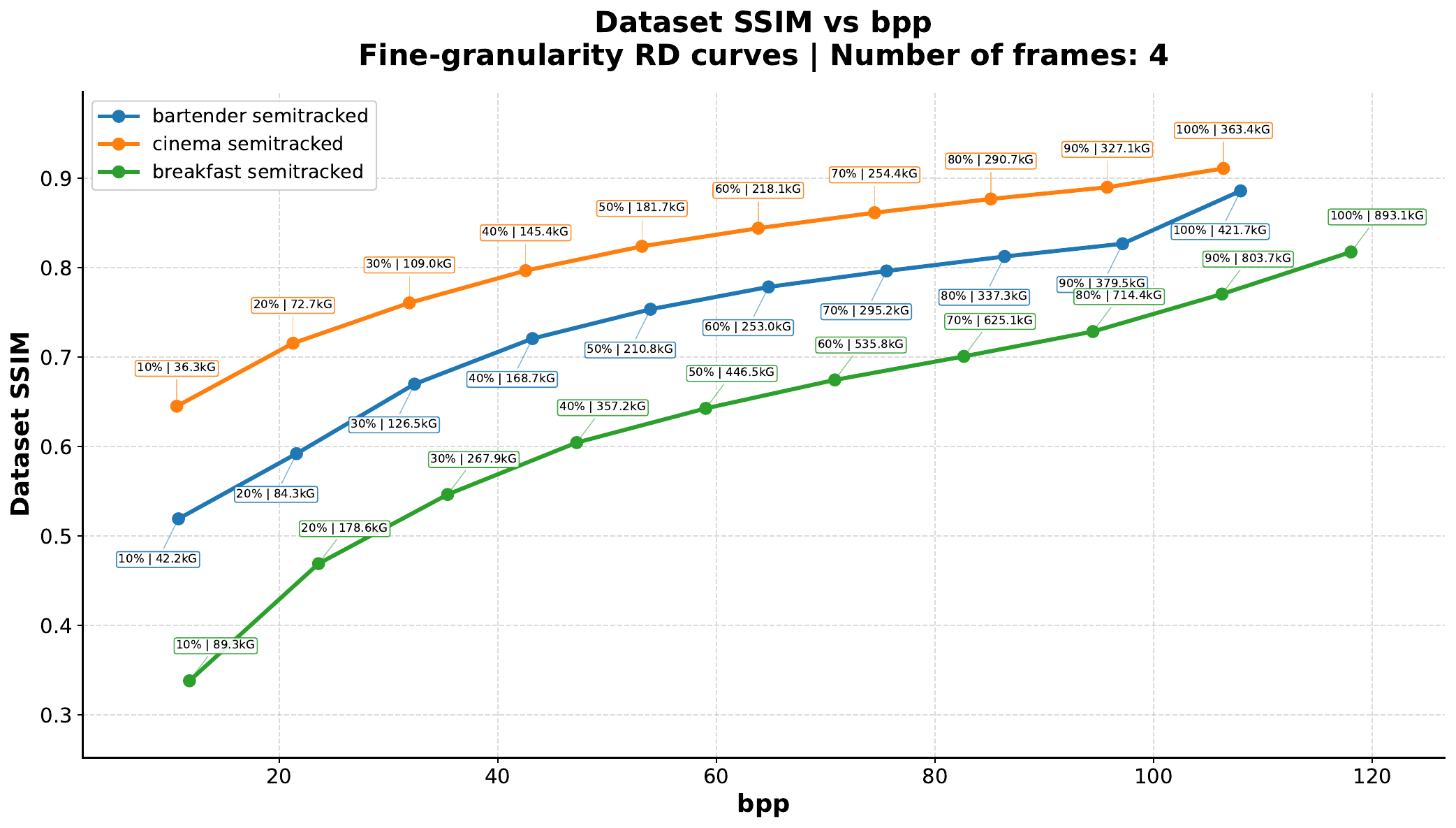}
    \caption{SSIM versus bpp for fine-grained refinement levels across three dynamic 3DGS sequences.}
    \label{fig:rd_ssim}
    \vspace{-5mm}
\end{figure}

\begin{figure}[t]
    \centering
    \includegraphics[width=\linewidth]{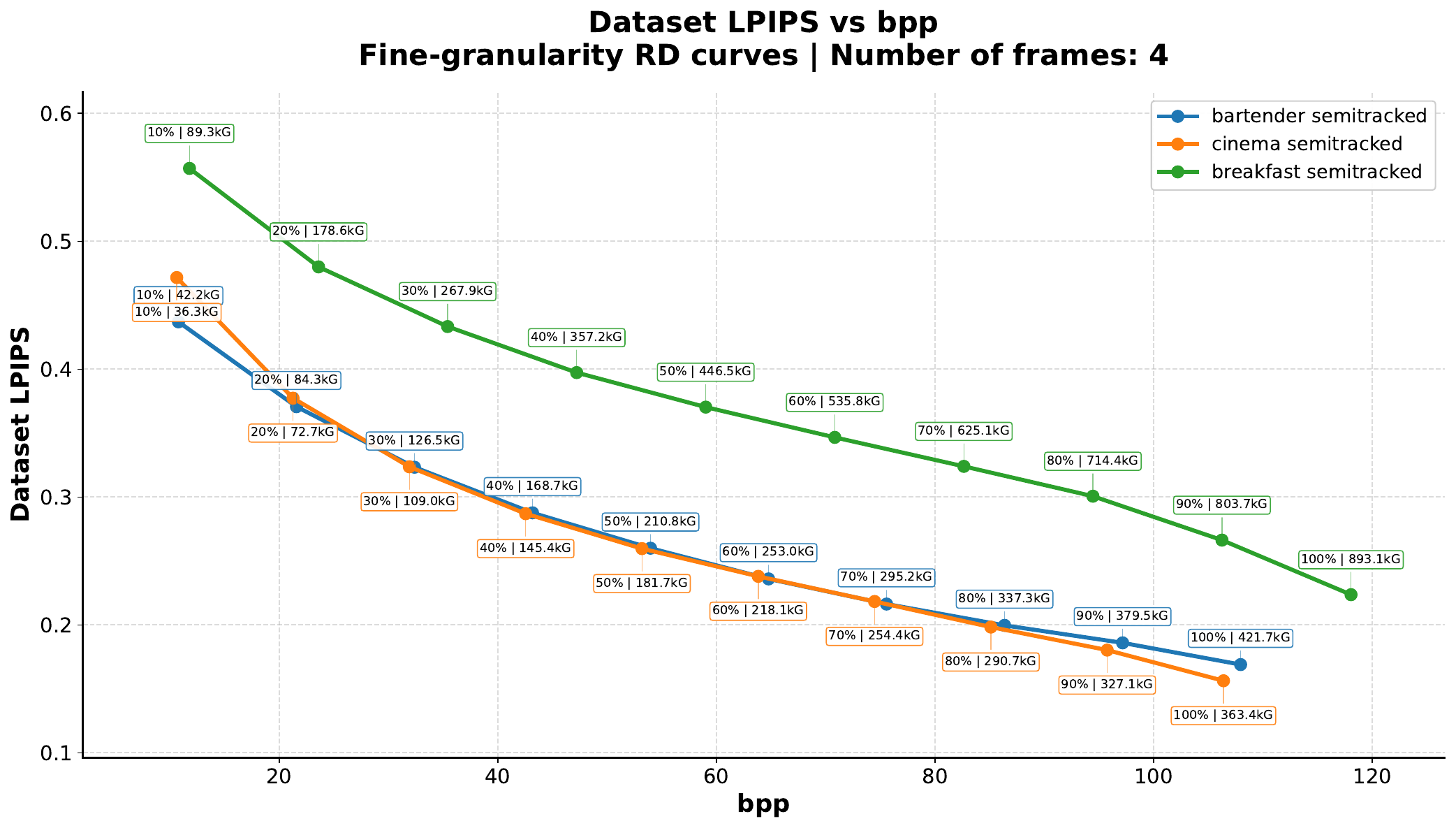}
    \caption{LPIPS versus bpp for fine-grained refinement levels across three dynamic 3DGS sequences.}
    \label{fig:rd_lpips}
    \vspace{-5mm}
\end{figure}

\vspace{-2mm}
\subsection{Transformer-Enhanced InterGS-Lite Prediction}
Temporal prediction in SplatStream is built upon InterGS-Lite~\cite{talha2026_intergslite}. The original InterGS-Lite framework uses a hybrid predictor based on K-nearest-neighbor (KNN) feature transfer and bilateral prediction. In this work, we extend the predictor set by adding a lightweight transformer predictor. Thus, for each target Gaussian, the encoder considers multiple prediction candidates:
\vspace{-2mm}
\begin{equation}
    \mathcal{P}
    =
    \{
    \mathcal{P}_{\text{KNN}},
    \mathcal{P}_{\text{bilateral}},
    \mathcal{P}_{\text{trans}}
    \}.
    \vspace{-1mm}
\end{equation}
The KNN branch captures local geometric correspondence from the reconstructed reference frame, the bilateral branch provides a spatially smooth attribute prediction, and the transformer branch learns geometry-aware local correlations that are difficult to capture with hand-crafted predictors.

For a P-frame at time $t$, prediction is performed in the reconstructed $1080p$ Gaussian domain. Given the previous reconstructed reference frame $\tilde{\mathcal{G}}_{t-1}^{1080}$ and the current target frame $\mathcal{G}_{t}^{1080}$, each predictor estimates the attributes of a target Gaussian:
\vspace{-2mm}
\begin{equation}
    \hat{\boldsymbol{\theta}}_{t,i}^{(p)}
    =
    \mathcal{P}_{p}
    \left(
    g_{t,i},
    \tilde{\mathcal{G}}_{t-1}^{1080}
    \right),
    \quad
    p\in\{\text{KNN},\text{bilateral},\text{trans}\}.
\end{equation}
The encoder selects the predictor that minimizes the attribute prediction error:
\vspace{-3mm}
\begin{equation}
    p_i^{*}
    =
    \arg\min_{p}
    \left\|
    \boldsymbol{\theta}_{t,i}
    -
    \hat{\boldsymbol{\theta}}_{t,i}^{(p)}
    \right\|_1 .
\end{equation}
The selected prediction is used to form the residual:
\vspace{-2mm}

\begin{equation}
    \mathbf{e}_{t,i}
    =
    \boldsymbol{\theta}_{t,i}
    -
    \hat{\boldsymbol{\theta}}_{t,i}^{(p_i^{*})}.
    \vspace{-2mm}
\end{equation}
The predictor index $p_i^{*}$ is transmitted as side information so that the decoder can reproduce the same prediction.

For the transformer branch, the input is formed from a local KNN neighborhood in the reconstructed reference frame. For each target Gaussian $g_{t,i}$, we collect $K$ neighboring reference Gaussians:
\vspace{-2mm}
\begin{equation}
    \mathcal{N}_{i}
    =
    \text{KNN}
    \left(
    \mathbf{x}_{t,i},
    \tilde{\mathcal{G}}_{t-1}^{1080}
    \right).
    \vspace{-2mm}
\end{equation}
Each neighbor is converted into a token containing compact PCA-domain SH information and relative geometry. Let $\boldsymbol{\tau}_{ij}$ denote the token of the $j$-th neighbor and $w_{ij}$ its affinity weight. The token is first affinity-gated:
\begin{equation}
    \mathbf{u}_{ij}
    =
    w_{ij}\boldsymbol{\tau}_{ij}.
\end{equation}
The gated token is projected into an embedding space and augmented with learned rank and geometric positional information:
\begin{equation}
    \mathbf{h}_{ij}^{0}
    =
    W_{\text{in}}\mathbf{u}_{ij}
    +
    \mathbf{r}_{j}
    +
    f_{\text{geo}}(\Delta\mathbf{x}_{ij}),
    \vspace{-2mm}
\end{equation}
where $\mathbf{r}_{j}$ is a learned rank embedding, $f_{\text{geo}}(\cdot)$ is a geometry MLP, and $\Delta\mathbf{x}_{ij}$ is the relative position between the target Gaussian and the reference neighbor.

The transformer applies single-head QKV attention:
\begin{equation}
    \mathbf{Q}_{i}=\mathbf{H}_{i}^{0}W_Q,\quad
    \mathbf{K}_{i}=\mathbf{H}_{i}^{0}W_K,\quad
    \mathbf{V}_{i}=\mathbf{H}_{i}^{0}W_V.
\end{equation}
To make the attention explicitly aware of local 3D geometry, a relative-position bias is added to the attention logits:
\vspace{-2mm}
\begin{equation}
    S_i(m,n)
    =
    \frac{\mathbf{q}_{im}^{T}\mathbf{k}_{in}}{\sqrt{d}}
    +
    b_{\psi}
    \left(
    \Delta\mathbf{x}_{in}
    -
    \Delta\mathbf{x}_{im}
    \right),
    \vspace{-2mm}
\end{equation}
where $b_{\psi}(\cdot)$ is implemented using Fourier relative-position features followed by a learned projection. The attention weights and context are computed as
\vspace{-2mm}
\begin{equation}
    \mathbf{A}_{i}
    =
    \text{softmax}(\mathbf{S}_{i}),
    \qquad
    \mathbf{C}_{i}
    =
    \mathbf{A}_{i}\mathbf{V}_{i}.
    \vspace{-2mm}
\end{equation}
After residual connections, layer normalization, and a feed-forward block, the predictor head outputs the PCA-domain SH prediction:
\vspace{-3mm}
\begin{equation}
    \hat{\mathbf{y}}_{t,i}^{\text{trans}}
    =
    f_{\text{head}}
    \left(
    \text{Pool}(\mathbf{H}_{i})
    \right).
    \vspace{-2mm}
\end{equation}
The transformer is trained using an L1 loss against the target PCA-domain SH vector:
\vspace{-2mm}
\begin{equation}
    \mathcal{L}_{\text{trans}}
    =
    \frac{1}{N}
    \sum_{i=1}^{N}
    \left\|
    \hat{\mathbf{y}}_{t,i}^{\text{trans}}
    -
    \mathbf{y}_{t,i}
    \right\|_1.
\end{equation}

After predictor selection, SH residuals are coded using vector quantization, while opacity, scale, and rotation residuals are scalar-quantized and entropy coded, following InterGS-Lite. At the decoder, the same reconstructed reference frame and transmitted predictor index are used to reproduce the selected prediction. The decoded frame is reconstructed as
\begin{equation}
    \tilde{\mathcal{G}}_{t}^{1080}
    =
    \hat{\mathcal{G}}_{t}^{1080}
    +
    \tilde{\mathcal{E}}_{t}^{1080}.
\end{equation}
Thus, SplatStream preserves the closed-loop InterGS-Lite prediction structure while extending its predictor set with a learned geometry-aware transformer.

\begin{table*}[t]
\caption{Fine-grained importance-ordered refinement operating points after the full intra anchor. The $P_2$, $B_1$, and $B_3$ columns report the coded size of the selected refinement unit for display frames 2, 1, and 3, respectively. Added temporal size is their sum. Total size includes the full intra anchor S3. Total Gaussians and cumulative BPG are computed over frames 0, 1, 2, and 3, and quality metrics are averaged over these frames.}
\label{tab:fine_grained_refinement}
\vspace{-1mm}
\centering
\scriptsize
\resizebox{\textwidth}{!}{
\begin{tabular}{c c c c c c c c c c}
\hline
\textbf{Ref. level} &
\textbf{Coding level} &
\textbf{$P_2$ size} &
\textbf{$B_1$ size} &
\textbf{$B_3$ size} &
\textbf{Added temp. size} &
\textbf{Total size} &
\textbf{Total \#G} &
\textbf{Cum. BPG} &
\textbf{Avg. PSNR / SSIM / LPIPS} \\
\hline
10\% &
Very-low-rate ref. &
0.59 MB &
0.53 MB &
0.59 MB &
1.71 MB &
23.46 MB &
474560 &
395.48 &
22.86 / 0.617 / 0.365 \\
\hline
20\% &
Low-rate ref. &
1.17 MB &
1.06 MB &
1.18 MB &
3.41 MB &
25.17 MB &
601059 &
334.97 &
24.51 / 0.672 / 0.316 \\
\hline
30\% &
Progressive ref. &
1.76 MB &
1.60 MB &
1.76 MB &
5.12 MB &
26.87 MB &
727557 &
295.49 &
25.92 / 0.730 / 0.280 \\
\hline
50\% &
Medium-rate ref. &
2.93 MB &
2.66 MB &
2.94 MB &
8.53 MB &
30.29 MB &
980553 &
247.10 &
27.82 / 0.793 / 0.233 \\
\hline
100\% &
Full-quality ref. &
5.86 MB &
5.32 MB &
5.88 MB &
17.07 MB &
38.82 MB &
1613044 &
192.53 &
32.34 / 0.892 / 0.164 \\
\hline
\end{tabular}
}
\vspace{-7mm}
\end{table*}
\vspace{-2mm}
\subsection{Hierarchical GOP and B-Layer Temporal Scalability}

To support temporal scalability, SplatStream organizes each GOP into a base temporal layer and an enhancement temporal layer. For example, for a seven-frame GOP,
    $\{1,2,3,4,5,6,7\},$
the base temporal layer contains $I_1,\;P_3,\;P_5,\;P_7,$
which provides a lower-frame-rate representation. The base temporal prediction chain is
\vspace{-3mm}
\begin{equation}
    I_1 \rightarrow P_3 \rightarrow P_5 \rightarrow P_7.
    \vspace{-2mm}
\end{equation}
The intermediate frames are assigned to a B-layer temporal enhancement representation:
  $  B_2,\;B_4,\;B_6. $
For simplicity, these B-layer frames are not coded with true bidirectional prediction. Instead, they are encoded using the same InterGS-Lite P-style prediction from the nearest previous reconstructed anchor:
\vspace{-2mm}
\begin{equation}
    I_1\rightarrow B_2,\quad
    P_3\rightarrow B_4,\quad
    P_5\rightarrow B_6.
    \vspace{-2mm}
\end{equation}
Thus, the term B-layer denotes the temporal enhancement role of these frames, not a bidirectional prediction mode. A client may decode only the base temporal layer for lower-frame-rate playback or request B-layer frames to recover the full frame rate.

\subsection{Opacity-Volume Prefixes for Fine-Grained Scalability}

SplatStream further supports fine-grained quality scalability within each decoded Gaussian frame. After a frame is reconstructed, its Gaussians are ranked using an opacity-volume importance score. Since 3DGS stores opacity as a logit and scale in log-space, we compute
\vspace{-2mm}
\begin{equation}
    V_i=\exp(s_{i,x}+s_{i,y}+s_{i,z}),
    \vspace{-2mm}
\end{equation}
\begin{equation}
    \rho_i=\sigma(\alpha_i)V_i,
    \vspace{-2mm}
\end{equation}
where $\sigma(\cdot)$ is the sigmoid function and $\rho_i$ is the importance score. Gaussians with larger spatial support and higher opacity are transmitted earlier.

Let $\pi$ be the ordering that sorts Gaussians by decreasing $\rho_i$. For a prefix percentage $x$, the renderable subset is
\begin{equation}
    \tilde{\mathcal{G}}_{t}^{x}
    =
    \left\{
    g_{\pi(i)}
    \mid
    1\leq i \leq
    \left\lfloor
    \frac{xN_t}{100}
    \right\rfloor
    \right\},
    \vspace{-2mm}
\end{equation}
where $N_t$ is the number of Gaussians in the reconstructed frame. Prefixes form nested representations:
\vspace{-2mm}
\begin{equation}
    \tilde{\mathcal{G}}_{t}^{10}
    \subset
    \tilde{\mathcal{G}}_{t}^{20}
    \subset
    \cdots
    \subset
    \tilde{\mathcal{G}}_{t}^{100}.
    \vspace{-2mm}
\end{equation}
This enables progressive rendering from a small subset of important Gaussians to the full reconstruction.
\vspace{-1mm}

\begin{table*}[t]
\caption{Bandwidth-adaptive operating points of SplatStream for one four-frame GOP. Total size is the transmitted data required for the selected operating point. Required bandwidth is computed assuming a 30 fps stream and a four-frame GOP. Quality metrics are averaged over the available decoded frames.}
\vspace{-1mm}
\label{tab:bandwidth_operating_points}
\centering
\scriptsize
\resizebox{\textwidth}{!}{
\begin{tabular}{c c c c c c c c c}
\hline
\textbf{Bandwidth mode} &
\textbf{Selected stream units} &
\textbf{Available frames} &
\textbf{Playback mode} &
\textbf{Refinement} &
\textbf{Total size/GOP} &
\textbf{Req. BW} &
\textbf{Total \#G} &
\textbf{Avg. PSNR / SSIM / LPIPS} \\
\hline
Startup &
$I_{0}^{540}$ &
$0$ &
Intra only &
-- &
13.80 MB &
828 Mbps &
278127 &
30.45 / 0.886 / 0.161 \\
\hline
Low spatial quality &
$I_{0}^{540}+I_{0}^{720}$ &
$0$ &
Intra only &
-- &
17.56 MB &
1054 Mbps &
292580 &
33.15 / 0.904 / 0.155 \\
\hline
Full spatial anchor &
$I_{0}^{540}+I_{0}^{720}+I_{0}^{1080}$ &
$0$ &
Intra only &
-- &
21.75 MB &
1305 Mbps &
348060 &
34.12 / 0.911 / 0.150 \\
\hline
Low-frame-rate temporal &
$S3+P_{2}^{1080}$ &
$0,2$ &
Base temporal layer &
100\% &
27.62 MB &
1657 Mbps &
772201 &
32.84 / 0.898 / 0.160 \\
\hline
Very-low BW full FPS &
$S3+P_{2}^{10}+B_{1}^{10}+B_{3}^{10}$ &
$0,1,2,3$ &
Full temporal layer &
10\% &
23.46 MB &
1408 Mbps &
474560 &
22.86 / 0.617 / 0.365 \\
\hline
Low BW full FPS &
$S3+P_{2}^{20}+B_{1}^{20}+B_{3}^{20}$ &
$0,1,2,3$ &
Full temporal layer &
20\% &
25.17 MB &
1510 Mbps &
601059 &
24.51 / 0.672 / 0.316 \\
\hline
Moderate BW full FPS &
$S3+P_{2}^{30}+B_{1}^{30}+B_{3}^{30}$ &
$0,1,2,3$ &
Full temporal layer &
30\% &
26.87 MB &
1612 Mbps &
727557 &
25.92 / 0.730 / 0.280 \\
\hline
Medium BW full FPS &
$S3+P_{2}^{50}+B_{1}^{50}+B_{3}^{50}$ &
$0,1,2,3$ &
Full temporal layer &
50\% &
30.29 MB &
1817 Mbps &
980553 &
27.82 / 0.793 / 0.233 \\
\hline
High BW full FPS &
$S3+P_{2}^{100}+B_{1}^{100}+B_{3}^{100}$ &
$0,1,2,3$ &
Full temporal layer &
100\% &
38.82 MB &
2329 Mbps &
1613044 &
32.34 / 0.892 / 0.164 \\
\hline
\end{tabular}
}
\vspace{-6mm}
\end{table*}

\vspace{-1mm}
\subsection{MPEG-DASH Sub-Representation Mapping}
SplatStream maps the scalable Gaussian bitstream to a DASH-like structure, where each GOP is treated as a temporal segment and spatial, temporal, and fine-grained quality layers are exposed as selectable representations or sub-representations. The scalable intra packets support startup and spatial-quality refinement, the base temporal layer enables lower-frame-rate playback, and B-layer packets provide optional temporal enhancement for full-frame-rate playback.

Importance-ordered Gaussian refinement units are exposed as fine-grained sub-representations, allowing the client to request partial refinement levels, e.g., 10\%, 20\%, 30\%, 50\%, or 100\%, according to bandwidth and rendering budget. The DASH-like manifest records packet type, temporal layer, refinement percentage, byte size, bits per Gaussian, Gaussian count, predictor metadata, and dependency information, enabling the client to select the appropriate operating point for its bandwidth, frame-rate, and quality requirements.

\vspace{-2mm}
\subsection{Adaptive Decoding and Playback}
\vspace{-1mm}
At the decoder, reconstruction follows the dependency graph specified in the manifest. A low-bandwidth client may first decode a compact intra representation for fast startup. To enter the full-resolution temporal prediction chain, the client decodes the full-quality intra anchor and reconstructs $\tilde{\mathcal{G}}_{1}^{1080}$. Subsequent anchor P-frames are decoded using the transformer-enhanced InterGS-Lite prediction chain. If bandwidth is limited, the client may decode only the base temporal layer:
    $I_1,\;P_3,\;P_5,\;P_7.$
If additional bandwidth is available, the client requests B-layer enhancement frames:
  $  B_2,\;B_4,\;B_6. $
Within each decoded frame, the client may render only the available opacity-volume prefix $\tilde{\mathcal{G}}_{t}^{x}$, where $x\in\{10,20,\ldots,100\}$. As more sub-representations arrive, the rendering is progressively refined in quality and temporal smoothness.

Overall, SplatStream combines multi-resolution intra rendering supervision, transformer-enhanced InterGS-Lite prediction, B-layer temporal scalability, opacity-volume Gaussian prefixing, and MPEG-DASH sub-representation mapping. This unified design enables adaptive dynamic 3DGS streaming across display quality, frame rate, and fine-grained reconstruction levels.

\vspace{-3mm}
\section{Experimentation and Results}

We evaluate SplatStream on three forward-facing dynamic 3DGS sequences from the MPEG common test conditions: \textit{bartender\_semitracked}, \textit{cinema\_semitracked}, and \textit{breakfast\_semitracked}. Each experiment uses a four-frame GOP with display frames $\{0,1,2,3\}$, organized as $I_0,B_1,P_2,B_3$. Here, $I_0$ is the scalable intra anchor, $P_2$ forms the base temporal layer, and $B_1$ and $B_3$ are B-layer temporal enhancement frames coded using P-style prediction. The intra anchor supports multiple spatial qualities, while temporal prediction is performed in the reconstructed 1080p Gaussian domain. For fine-grained streaming, reconstructed Gaussians are ordered using the opacity-volume importance score and progressively decoded using refinement levels from 10\% to 100\%. We report RGB-PSNR, SSIM, and LPIPS. Required bandwidth is computed assuming a 30 fps stream and a four-frame GOP:
\vspace{-2mm}
\begin{equation}
    R_{\text{bw}}
    =
    \frac{8B_{\text{GOP}}}{T_{\text{GOP}}},
    \quad
    T_{\text{GOP}}=\frac{4}{30},
    \vspace{-2mm}
\end{equation}
where $B_{\text{GOP}}$ is the transmitted size of the selected operating point. Table~\ref{tab:bandwidth_operating_points} reports bandwidth-adaptive operating points for one GOP of the \textit{bartender\_semitracked} sequence. The first three rows show intra-only spatial scalability, where the client starts from $I_0^{540}$ and progressively receives $I_0^{720}$ and $I_0^{1080}$. The total size increases from 13.80 MB to 21.75 MB, while PSNR improves from 30.45 dB to 34.12 dB. After the full spatial anchor is available, the stream enters temporal playback. The base temporal mode adds $P_2^{1080}$ and enables frames $\{0,2\}$. Full-frame-rate playback is then achieved by adding refinement units for $P_2$, $B_1$, and $B_3$. As the refinement level increases from 10\% to 100\%, the total size increases from 23.46 MB to 38.82 MB, while PSNR improves from 22.86 dB to 32.34 dB and LPIPS decreases from 0.365 to 0.164. This demonstrates that SplatStream provides selectable DASH-like operating points under different bandwidth conditions.

Table~\ref{tab:fine_grained_refinement} further analyzes the fine-grained refinement units. The added temporal size increases smoothly from 1.71 MB at 10\% refinement to 17.07 MB at 100\% refinement. Over the same range, the number of rendered Gaussians increases from 474,560 to 1,613,044, PSNR improves from 22.86 dB to 32.34 dB, SSIM increases from 0.617 to 0.892, and LPIPS decreases from 0.365 to 0.164. These results show that opacity-volume ordering produces meaningful progressive refinement: early units provide a compact coarse reconstruction, while later units recover finer details and improve perceptual quality.

Fig.~\ref{fig:rd_psnr}, Fig.~\ref{fig:rd_ssim}, and Fig.~\ref{fig:rd_lpips} report fine-granularity RD curves over all three sequences. Across \textit{bartender\_semitracked}, \textit{cinema\_semitracked}, and \textit{breakfast\_semitracked}, increasing the refinement percentage consistently improves PSNR, SSIM, and LPIPS. Although the absolute quality varies by sequence due to different scene complexity and Gaussian distributions, all curves follow the same trend. This confirms that the proposed importance-ordered refinement units provide stable bandwidth-quality scalability across dynamic 3DGS content.

\vspace{-2mm}
\section{Conclusion}
\vspace{-1mm}
This paper presented \textit{SplatStream}, a fine-grained scalable streaming framework for dynamic 3D Gaussian Splatting. The proposed method converts dynamic 3DGS from a monolithic compressed representation into a DASH-like adaptive stream with spatial, temporal, and fine-grained quality scalability. Multi-resolution rendering supervision enables a compact intra anchor for startup and progressive spatial refinement, while the transformer-enhanced InterGS-Lite predictor improves temporal coding for both base P-frames and B-layer enhancement frames. In addition, opacity-volume based Gaussian ordering provides progressively decodable refinement units, allowing clients to trade bandwidth for reconstruction quality in a controlled manner. Experimental results on dynamic 3DGS sequences show that SplatStream supports bandwidth-adaptive operation, low-frame-rate and full-frame-rate playback, and smooth quality refinement across PSNR, SSIM, and LPIPS. These results demonstrate the potential of scalable Gaussian packetization and DASH-compatible organization for practical low-latency immersive 3D scene streaming.

\bibliographystyle{IEEEtran}
\bibliography{refs}

\end{document}